\documentclass[11pt]{article} 
\usepackage{a4wide}  
\RequirePackage[latin1]{inputenc}     
 
\usepackage[english]{babel} 
\usepackage{amsmath} 
\usepackage{amsfonts} 
\usepackage{amssymb} 
\usepackage{xspace} 
\usepackage{latexsym} 
\usepackage{url} 
\usepackage{xspace} 
\usepackage[all]{xy} 

\usepackage{graphics,color}              
\RequirePackage[latin1]{inputenc}

\newenvironment{restate-proposition}[2][{}]{\noindent\textbf{Proposition~{#2}}\;\textbf{#1}\  
}{\vskip 1em} 
 
\newenvironment{restate-theorem}[2][{}]{\noindent\textbf{Theorem~{#2}}\;\textbf{#1}\  
}{\vskip 1em} 
 
\newenvironment{restate-corollary}[2][{}]{\noindent\textbf{Corollary~{#2}}\;\textbf{#1}\  
}{\vskip 1em}

\newcommand{\Proofitem}[1]{\medskip \noindent $#1\;$} 
\newcommand{\Proofitemf}[1]{\noindent $#1\;$} 
\newcommand{\Defitem}[1]{\smallskip \noindent $#1\;$} 
 
\newcommand{\Defitemf}[1]{\noindent $#1\;$}

\newcommand{\hbra}{\noindent\hbox to \textwidth{\leaders\hrule height1.8mm depth-1.5mm\hfill}} 
\newcommand{\hket}{\noindent\hbox to \textwidth{\leaders\hrule height0.3mm\hfill}} 
\newcommand{\ratio}{.3}

\newtheorem{theorem}{Theorem} 
 
\newtheorem{definition}[theorem]{Definition} 
 
\newtheorem{corollary}[theorem]{Corollary} 
\newtheorem{proposition}[theorem]{Proposition}

\newtheorem{remark}[theorem]{Remark}

\newcommand{\Proof}{\noindent {\sc Proof}. } 
\newcommand{\Proofhint}{\noindent {\sc Proof hint}. } 
\newcommand{\qed}{\hfill${\Box}$}

\newcommand{\Figbar}{{\center \rule{\hsize}{0.3mm}}}    
 
\newcommand{\cl}[1]{{\cal #1}}          

\newcommand{\Gives}{\vdash}             
\newcommand{\Models}{\mid \! =}              

\newcommand{\impl}{\supset} 
\newcommand{\arrow}{\rightarrow}        
\newcommand{\Alt}{ \mid\!\!\mid  }

\newcommand{\infer}[2]{\begin{array}{c} #1 \\ \hline #2 \end{array}} 
 
\newcommand{\AND}{\wedge}               
 
\newcommand{\union}{\cup}               
\newcommand{\minus}{\backslash}         
\newcommand{\set}[1]{\{#1\}}            
 
\newcommand{\dcl}{\downarrow}           
\newcommand{\Conv}{\Downarrow}          

\newcommand{\w}[1]{{\it #1}}    

\newcommand{\qqs}[2]{\forall\, #1\;\: #2}

\newcommand{\s}[1]{{\sf #1}}    
\newcommand{\vc}[1]{{\bf #1}}

\newcommand{\sem}[1]{\underline{#1}} 
\newcommand{\tra}[1]{\langle #1 \rangle}

\newcommand{\act}[1]{\xrightarrow{#1}} 

 \newcommand{\pres}[2]{#1\triangleright #2} 

\newcommand{\tick}{{\sf tick}}          

\newcommand{\regterm}[2]{{\sf reg}_{#1} #2}

\newcommand{\store}[2]{(#1 \Leftarrow #2)}
\newcommand{\gt}[1]{{\sf get}(#1)}
\newcommand{\st}[2]{{\sf set}(#1,#2)}
\newcommand{\regtype}[2]{{\sf Reg}_{#1} #2}

\newcommand{\Beh}{{\bf B}}
\newcommand{\Ter}{{\bf 1}}

\begin{document}

\title{On stratified regions}

\author{Roberto M. Amadio \\
        Universit\'e Paris Diderot (Paris 7)\thanks{PPS, UMR-CNRS 7126. 
Work partially supported by ANR-06-SETI-010-02.}}

\maketitle 

\begin{abstract}
Type and effect systems are a tool to analyse statically
the behaviour of programs with effects.
We present a proof 
based on the so called reducibility candidates 
that a suitable stratification of the type and
effect system entails the termination of the typable programs.
The proof technique covers a simply typed, multi-threaded, call-by-value lambda-calculus,
equipped with a variety of
scheduling (preemptive, cooperative) and interaction mechanisms 
(references, channels, signals). \\
\noindent {\bf Keywords} Types and effects. Termination. Reducibility candidates.
\end{abstract}

\section{Introduction}
In the framework of functional programs, the relationship between type
systems and termination has been extensively studied through the
Curry-Howard correspondence.  It would be interesting to extend these
techniques to programs with {\em effects}. By effect we mean the
possibility of executing operations that modify the state of a system
such as reading/writing a reference or sending/receiving a message.

Usual type systems as available, {\em e.g.}, in various dialects
of the {\sc ML} programming language, are too poor to account for the behaviour
of programs with effects. A better approximation
is possible if one abstracts the state of a system in a certain
number of {\em regions} and if the types account for the way programs
act on such regions.  So-called {\em type and effect} systems \cite{LG88} are
an interesting formalisation of this idea and have been 
successfully used to analyse statically the problem of heap-memory
deallocation \cite{TT97}. 
On the other hand, the proof-theoretic foundations of such systems
are largely unexplored. Only recently, it has been shown \cite{Boudol07}
that a {\em stratification} of the regions entails
termination in a certain higher-order language with cooperative threads and 
references. Our purpose here is to revisit this result trying to 
clarify and extend both its scope and its proof technique 
(a more technical comparison is delayed to section \ref{stratified}).
We refer to \cite{Boudol07} for a tentative list of papers referring
to a notion of stratification for programs with side effects. 
Perhaps the closest works in spirit are those that have adapted
the reducibility candidates techniques to the $\pi$-calculus  
\cite{YBH04,S06}.
Those works exhibit type systems for the $\pi$-calculus
that guarantee the termination of the usual continuation passing 
style translations of typed {\em functional} languages into the
$\pi$-calculus. However, as pointed out by one of the  authors of {\em
op.cit} in \cite{DS06},  they are not very  successful in 
handling state sensitive programs.
The approach here is a bit different: one starts with a 
higher-order typed functional language which is known
to be terminating and then one determines to what extent 
side-effects can be added while preserving termination.
Yet in another direction, we notice that a notion of region stratification
has been used in \cite{AD07} to guarantee the polynomial time reactivity
of a first-order timed/synchronous language.

We outline the contents of the paper.
In section \ref{lambda-region}, we introduce a $\lambda$-calculus with
{\em regions}.  Regions are an abstraction of dynamically generated
values such as references, channels, and signals, and the reduction
rules of the calculus are given in such a way that the reduction rules
for references, channels, and signals can be simulated by those 
given for regions.
In section \ref{unstratified}, we describe a simple {\em type and
effect} system along the lines of \cite{LG88}. In this discipline,
types carry information on the regions on which the evaluated
expressions may read or write.  The discipline allows to write in a
region $r$ values that have an effect on the region $r$
itself. In turn, this allows to simulate recursive definitions and
thus to produce non terminating behaviours.
In section \ref{stratified}, following  \cite{Boudol07}, 
we describe a stratification of the regions. The idea is that regions are
ordered and that a value written in a region may only produce effects
in smaller regions. We then propose a new  reducibility candidates 
interpretation (see, {\em e.g.}, \cite{Gal90} for a good survey)
entailing the termination of typable programs.
In section \ref{extension-sec}, we enrich the language 
with the possibility to generate new threads and to
react to the termination of the computation.  
The language we consider is then {\em
timed/synchronous} in the sense that a computation is regarded as a possibly
infinite sequence of instants.  An instant ends when the calculus
cannot progress anymore (cf. timed/synchronous languages such as
Timed CCS \cite{HR95} and {\sc Esterel} \cite{BG92}).  
We extend the stratified typing rules to this language and
show by means of a translation into the core language that typable
programs terminate.  We also show that a fixed-point combinator can
be {\em defined} and {\em typed} so that recursive calls are allowed
as long as they arise at a later instant. This differs from \cite{Boudol07}
where a fixed-point combinator is {\em added} to the language potentially
compromising the termination property.
Appendix \ref{proofs-sec} contains the main proofs and appendix \ref{summary-sec}
summarizes the type and effect systems considered.

\section{A $\lambda$-calculus with regions}\label{lambda-region}
We consider a $\lambda$-calculus with {\em regions}.  Regions are {\em
abstractions} of dynamically generated `pointers' which, depending on
the context, are called references, channels, or signals.  Given a
program with operators to generate dynamically values (such as \s{ref}
in the ML language or $\nu$ in the $\pi$-calculus), one may simply
introduce a distinct region for every occurrence of such operators.
This amounts to collapse all the `pointers' generated by the operator
at run time into one constant. The resulting language simulates the
original one as long as the values written into regions do not erase
those already there. In particular, termination for the language with
regions entails termination for the original language.

We notice that ordinary type system for programs with dynamic values 
perform  a similar abstraction: all the values that are generated by an operator
are assigned the same type. For instance, typing $\nu x \ P$ in the
$\pi$-calculus will reduce to typing the process $P$ 
in a context where the name $x$ is associated with a suitable type $A$.
In the corresponding language with regions, one will replace 
the name $x$ with a region $r$ and type $[r/x]P$ ($[r/x]$ is the substitution)
in a region context where $r$ is associated with $A$.

To summarise, termination for the language with regions entails
termination for the original calculi and moreover ordinary type system 
implicitly abstract dynamically generated values into regions.
Therefore, we argue that one can carry on the main type theoretic arguments at
the level of regions rather than at the more detailed level of dynamically
generated values. \footnote{Incidentally, it seems much easier 
to produce denotational models of languages with regions than for the original 
languages with dynamic values so that one can hope to find
models that do provide insight into the type systems.}

\subsection{Syntax}
We consider the following syntactic categories:
\[
\begin{array}{ll}

x,y,\ldots         
&\mbox{(variables)} \\

r,s,\ldots         
&\mbox{(regions)} \\

e,e',\ldots
&\mbox{(finite sets of regions)} \\

A::= \Ter \Alt \regtype{r}{A} \Alt (A \act{e} A)
&\mbox{(types)} \\

\Gamma::= x_1:A_1,\ldots,x_n:A_n
&\mbox{(context)} \\

R::= r_1:A_1,\ldots,r_n:A_n 
&\mbox{(region context)} \\

M::= x \Alt r \Alt * \Alt \lambda x.M \Alt MM \Alt \gt{M} \Alt \st{M}{M}  &\mbox{(terms)} \\

V::= r \Alt * \Alt \lambda x.M
&\mbox{(values)} \\

v,v',\ldots
&\mbox{(sets of value)} \\
 
S::= \store{r}{v} \Alt S,S 
&\mbox{(stores)} \\

X::= M \Alt S
&\mbox{(stores or terms)}\\

P::= X \Alt X,P 
&\mbox{(programs)}

\end{array}
\]
We briefly comment the notation:
$\Ter$ is the terminal (unit) type with value $*$;
$\regtype{r}{A}$ is the type of a region $r$ containing
values of type $A$;
$A\act{e}B$ is the type of functions that when given
a value of type $A$ may produce a value of type $B$
and an effect on the regions in $e$;
$\s{get}$ is the operator to read {\em some} value in  
a region and $\s{set}$ is the operator to {\em insert} a value in a region.

We write $[N/x]M$ for the substitution of $N$ for $x$ in $M$.
If $R=r_1:A_1,\ldots,r_n:A_n$ then $\w{dom}(R)=\set{r_1,\ldots,r_n}$.
If $r\in \w{dom}(R)$ then we write $R(r)$ for the type $A$ such that
$r:A$ occurs in $R$.
We also define the term $\regterm{r}{M}$ as an abbreviation for $(\lambda
x.r)(\st{r}{M})$. Thus the difference between $\st{r}{M}$ and $\regterm{r}{M}$
is that in the first case we return $*$ while in the second we return
$r$. 
When writing a program $P=X_1,\ldots,X_n$ we regard the symbol `,' as
associative and commutative, or equivalently we regard a program as a
multi-set of  terms and stores. 
We write 
$\store{r}{V}$ for $\store{r}{\set{V}}$. 
We shall identify the store $\store{r}{v_{1}}, \store{r}{v_{2}}$
with the store $\store{r}{v_{1}\union v_{2}}$.
We denote with $\w{dom}(S)$ the set of regions $r$ such that
$\store{r}{v}$ occurs in $S$ and define $S(r)$ as the set 
$\set{V\mid \store{r}{V} \mbox{ occurs in }S}$.

\subsection{Reduction}
A call-by value {\em evaluation context} $E$ is defined as:
\[
E::= [~] \Alt EM \Alt VE \Alt  \gt{E} \Alt \st{E}{M} \Alt \st{V}{E} 
\]
An {\em elementary} evaluation context is defined as:
\[
\w{El}::= [~]M \Alt V[~] \Alt  \gt{[~]} \Alt \st{[~]}{M} \Alt \st{V}{[~]}
\]
An evaluation context  can be regarded as the finite composition 
(possibly empty) of elementary evaluation contexts. 
The {\em reduction} on programs is defined as follows:
\[
\begin{array}{c}

\infer{}{E[(\lambda x.M)V] \arrow E[[V/x]M]} 
\qquad
\infer{}{E[\gt{r}],\store{r}{V} \arrow E[V],\store{r}{V}} \\ \\ 

\infer{}{E[\st{r}{V}]\arrow E[*],\store{r}{V}}

\qquad 
\infer{P\arrow P'}
{P,P'' \arrow P',P''}

\end{array}
\]
Note that the semantics of \s{set} amounts to {\em add} rather
than to {\em update} a binding between a region and a value. 
Hence a region can be bound at the same time to several values
(possibly infinitely many) and the semantics of \s{get}
amounts to select non-deterministically one of them.

As already mentioned, the notion of region is intended to 
simulate some familiar programming
concepts such as references, channels, or signals. 
Specifically: (i) when writing a reference, we replace the previously written
value (if any),  (ii) when reading a (unordered, unbounded) 
channel we consume (remove from the store)  the value read,
and finally (iii) the values written in a signal 
persist within an instant and disappear at the end of 
it.\footnote{Signals arise in timed/synchronous models where the computation is
regulated by a notion of instant or phase (see section \ref{extension-sec}).}
One can easily formalise the reduction rules for references, channels,
and signals, and check that (within an instant) each reduction step is
simulated by at least one reduction step in the calculus with regions.
Thus, typing disciplines that guarantee termination for the calculus
with regions will guarantee the same property when adapted to
references, channels, or signals.

\section{Types and effects: unstratified case}\label{unstratified}
We introduce  a simple {\em type and effect} system along the 
lines of \cite{LG88}.  
The following rules define when a region context $R$ 
is {\em compatible} with a type $A$ (judgement $R\dcl A$):
\[
\begin{array}{ccc}
\infer{}
{R\dcl \Ter}
\quad

&\infer{R\dcl A\quad R\dcl B \quad e\subseteq \w{dom}(R)}
{R\dcl (A\act{e} B)}

\quad

&\infer{r:A\in R}
{R\dcl \regtype{r}{A}}
\end{array}
\]
The compatibility relation is just introduced to 
define when a region context is well formed (judgement $R\Gives$)
and when a type and effect 
is well-formed with respect to a region context (judgements $R\Gives A$ and $R\Gives (A,e)$).
\[
\begin{array}{ccc}

\infer{\qqs{r\in \w{dom}(R)}{R\dcl R(r)}}
{R\Gives}

\quad
&\infer{R\Gives\quad R\dcl A}
{R\Gives A}

\quad
&\infer{R\Gives A\quad e\subseteq \w{dom}(R)}
{R\Gives (A,e)}
\end{array}
\]
A more informal way to express the condition is to say that a judgement
$r_1:A_1,\ldots,r_n:A_n \Gives B$ is well formed provided that:
(1) all the region names occurring in the types $A_1,\ldots,A_n,B$ belong to the
set $\set{r_1,\ldots,r_n}$ and (2) all types of the shape
$\regtype{r_{i}}{C}$ with $i\in \set{1,\ldots,n}$ and
occurring in the types $A_1,\ldots,A_n,B$ are such that $C=A_i$.
For instance, the reader may verify that 
$r:1\act{\set{r}} 1 \Gives \regtype{r}{1\act{\set{r}} 1}$ can be derived
while $r_1: \regtype{r_{2}}{(1\act{\set{r_{2}}} 1)}, r_2: 1\act{\set{r_{1}}} 1 \Gives$
cannot. Also it can be easily checked that the following properties hold:
\[
\begin{array}{lcl}

R\Gives \Ter &\mbox{iff} &R\Gives \\
R\Gives \regtype{r}{A} &\mbox{iff} &R\Gives \mbox{ and }R(r)=A \\
R\Gives A\act{e}B      &\mbox{iff} &R\Gives, R\Gives A, R\Gives B, \mbox{ and } e\subseteq \w{dom}(R) \\

R\Gives   &\mbox{iff} &\qqs{r\in \w{dom}(R)}{R\Gives R(r)} 
\end{array}
\]
The subset relation on effects induces a {\em subtyping} relation on 
types and on pairs of types and effects which is defined as follows (judgements 
$R\Gives A\leq A'$, $R\Gives (A,e)\leq (A',e')$):
\[
\begin{array}{c}

\infer{R\Gives A}{R\Gives A\leq A}

\qquad

\infer{
\begin{array}{c}
R\Gives A'\leq A\quad  R\Gives B\leq B'\\
e\subseteq e' \subseteq \w{dom}(R) 
\end{array}}
{R\Gives (A\act{e} B) \leq (A' \act{e'} B')} 

\qquad

\infer{\begin{array}{c}
R\Gives A \leq A'\\ 
e\subseteq e' \subseteq \w{dom}(R)
\end{array}}
{R\Gives (A,e) \leq (A',e')}

\end{array}
\]
We notice that the transitivity rule:
\[
\infer{R\Gives A \leq B\qquad R\Gives B\leq C}
{R\Gives A\leq C}
\]
can be derived via a simple induction on the height of the proofs.
The subtyping rule trades flexibility against precision of
the type system. For instance, suppose $A_1=\Ter\act{e_{1}}\Ter$ and
$A_2=\Ter \act{e_{2}} \Ter$ and we want to define the type $B$ of
the functionals that take a value $V_1$ of type $A_1$ and a value
$V_2$ of type $A_2$ and compute either $V_1 *$ or $V_2 *$.
We can define $B=A_1 \act{\emptyset} (A_2 \act{e_{1}\union e_{2}}
\Ter)$. The reader can check that both 
$\lambda x.\lambda y.x*$ and $\lambda x.\lambda y.y*$  have
type $B$ {\em provided} the subtyping rule is used. 
Incidentally, we note that \cite{Boudol07} seems to `forget' the
subtyping rule. While there are is no particular problems to provide a
reducibility candidates interpretation for this rule, we notice that
without it the following diverging ML expression 
$\s{let} \ l= \s{ref}(\lambda x.x)\ \s{in} \ l:=\lambda x.!lx; !l ()$, which is given
in {\em op.cit.} to motivate the stratification of regions does {\em not}
type already in the ordinary unstratified type and effect system 
because $(\lambda x.x)$ has type $\Ter\act{\emptyset} \Ter$ but not
$\Ter \act{\set{r}} \Ter$ where $r$ is the region associated with 
the reference $l$.

We now turn to the typing rules for the terms.
We shall write $R \Gives x_1:A_1,\ldots,x:A_n$ if $R\Gives$ and $R\Gives A_i$
for $i=1,\ldots,n$.  Note that in the following  rules 
we always refer to the {\em same} region context $R$.
\[
\begin{array}{c}

\infer{R\Gives \Gamma \quad x:A\in \Gamma}
{R;\Gamma\Gives x:(A,\emptyset)}

\qquad 
\infer{R\Gives \Gamma \quad r:A\in R}
{R;\Gamma\Gives r:(\regtype{r}{A},\emptyset)} 

\qquad
\infer{R\Gives \Gamma}
{R;\Gamma \Gives *: (\Ter,\emptyset)} \\ \\

\infer{R;\Gamma,x:A \Gives M: (B,e)}
{R;\Gamma\Gives \lambda x.M :(A\act{e} B,\emptyset)}

\qquad

\infer{R;\Gamma \Gives M: (A\act{e_{2}} B,e_{1}) \quad R;\Gamma \Gives N: (A,e_{3})}
{R;\Gamma \Gives MN :(B,e_1\union e_2\union e_3)} \\ \\

\infer{R;\Gamma \Gives M: (\regtype{r}{A}, e)}
{R;\Gamma \Gives \gt{M} : (A,e\union \set{r})} 

\qquad

\infer{R;\Gamma \Gives M: (\regtype{r}{A}, e_{1})\quad R;\Gamma \Gives N:(A,e_{2})}
{R;\Gamma \Gives \st{M}{N} : (\Ter,e_1\union e_2 \union \set{r})}  \\ \\ 

\infer{R;\Gamma \Gives M:(A,e) \quad R\Gives (A,e)\leq (A',e')}
{R;\Gamma \Gives M:(A',e')}

\end{array}
\]
Finally, we extend the typing rules to stores and general
multi-threaded programs. To this end, it is convenient to 
introduce a constant behaviour type $\Beh$ which is the 
type we give to multi-sets of threads and/or stores which
are not supposed to return a value but just to interact
via side-effects. We will use $\alpha,\alpha',\ldots$ to
denote either an ordinary type $A$ or this new behaviour type $\Beh$.
\[
\begin{array}{c}

\infer{r:A \in R \quad \qqs{V\in v}{R;\Gamma \Gives V: (A,\emptyset)}}
{R;\Gamma \Gives \store{r}{v} :(\Beh,\emptyset)}

\qquad 
\infer{R;\Gamma \Gives X_i:(\alpha_i,e_i)\quad i=1,\ldots,n\geq 1}
{R;\Gamma \Gives X_1,\ldots,X_n:(\Beh, e_1\union \cdots\union e_n)} 

\end{array}
\]

\begin{remark}
The derived typing rule for $\regterm{r}{M}$ is as follows:
\[
\infer{r:A\in R\quad R;\Gamma \Gives M : (A,e)}
{R;\Gamma \Gives \regterm{r}{M} : (\regtype{r}{A},e \union \set{r})}
\]
\end{remark}

One can derive a more traditional `effect-free' type system by {\em
erasing all the effects} from the types and the typing
judgements. Note that in the resulting system the subtyping rules are
useless.  We shall write $\Gives^{ef}$ for provability in this system.
This `weaker' type system suffices to state a decomposition property
of the terms which is proven by induction on the structure of the term.

\begin{proposition}[decomposition]\label{decomp}
If $R; \Gives^{\w{ef}} M: A$ is a well-typed closed term
then exactly one of the following situations arises where
$E$ is an evaluation context:
\begin{enumerate}

\item $M$ is a value.

\item $M=E[\Delta]$ and $\Delta$ has the shape $(\lambda x.N)V$, $\st{r}{V}$, or $\gt{r}$.

\end{enumerate}
\end{proposition}

\subsection{Basic properties of typing and evaluation}
We observe some basic properties: (i) one can weaken both 
the type and region contexts, (ii)  typing is preserved when 
we replace a variable with an effect-free term of the same type, and 
(iii) typing is preserved by reduction.
If $S$ is a store and $e$ is a set of regions then 
$S_{\mid e}$ is the store $S$ restricted to the regions in $e$.

\begin{proposition}[basic properties, unstratified]\label{basic-prop-unstratified}
The following properties hold:
\begin{description}

\item[weakening]
If $R;\Gamma \Gives M:(A,e)$ and $R,R' \Gives \Gamma,\Gamma'$ then
 $R,R';\Gamma,\Gamma' \Gives M:(A,e)$.

\item[substitution]
If $R;\Gamma,x:A \Gives M:(B,e)$ and 
$R;\Gamma \Gives N:(A,\emptyset)$ then 
$R;\Gamma \Gives [N/x]M:(B,e)$.

\item[subject reduction]
Let $\vc{M}$ denote a sequence $M_1,\ldots,M_n$.
If $R,R'; \Gives \vc{M},S:(\Beh,e)$, 
$R \Gives e$, and $\vc{M},S \arrow \vc{M'},S'$  then
$R,R'; \Gives \vc{M'},S':(\Beh,e)$,
$S_{\mid \w{dom}(R')} = S'_{\mid \w{dom}(R')}$, and
$\vc{M},S_{\mid \w{dom}(R)} \arrow \vc{M'},S'_{\mid\w{dom}(R)}$.
Moreover, if $\vc{M}=M$ and $R,R'\Gives M:(A,e)$ then 
$\vc{M'}=M'$ and  $R,R'\Gives M':(A,e)$.
\end{description}
\end{proposition}

The weakening and substitution properties are shown directly by
induction on the proof height. Concerning subject reduction, it is
useful to notice that if a term $M$, of type and effect $(A,e)$,
is ready to read/write the region $r$ then $r\in e$.  This follows
from an analysis of the evaluation context.  Then we prove the
assertion by case analysis on the reduction rule applied, relying on
the substitution property.

\begin{remark}\label{effect-delimitation-rmk}
The subject reduction property is formulated so as to make clear that
the type and effect system indeed delimits the interactions a term may
have with the store.  Note that a term may refer to regions which are
not explicitly mentioned in its type and effect. For instance,
consider $M=(\lambda f.*)(\lambda x.\gt{r} x)$ and let
$R=r: \Ter \act{\emptyset} \Ter $.  Then $R;\emptyset \Gives M:(\Ter,\emptyset)$,
$\emptyset \Gives (\Ter,\emptyset)$ but $\emptyset;\emptyset \not\Gives
M:(\Ter,\emptyset)$.  The subject reduction property guarantees that such
a term will only read/write regions included in the region context
needed to type its type and effect.
\end{remark}

\subsection{Recursion}
In our (unstratified) calculus, we can write in a
region $r$ a functional value $\lambda x.M$ where $M$ reads
from the region $r$ itself. For instance, 
$\regterm{r}{(\lambda x.(\gt{r})x)}$.

This kind of circularity leads to diverging computations such as:
\[
\begin{array}{llll}
\gt{\regterm{r}{\lambda x.\gt{r} x}}* 
&\arrow
&\gt{r}*, \store{r}{\lambda x.\gt{r} x} &\arrow \\

(\lambda x.\gt{r}x)*, \store{r}{\lambda x.\gt{r}x} 
&\arrow
&\gt{r}*, \store{r}{\lambda x.\gt{r}x} 
&\arrow \cdots
\end{array}
\]

It is well known that this phenomena can be exploited to 
simulate recursive definitions. Specifically, we define:
\begin{equation}\label{fix-def}
\s{fix}_{r} f.M = 
\lambda x.
(\gt{\regterm{r}{(\lambda x.[\lambda x.\gt{r}x/f]M\ x)}})\ x
\end{equation}
By a direct application of the typing rules and 
proposition \ref{basic-prop-unstratified}(substitution),
one can derive a  rule to type $\s{fix}_{r} f.M$.

\begin{proposition}[type fixed-point]\label{derived-rule1-fix}
The following typing rule for the fixed point combinator is derived:
\begin{equation}\label{fix-type-unstratified}
\infer{\begin{array}{c}
r:A\act{e}B \in R \quad r\in e \\
R;\Gamma,f:A\act{e}B\Gives M:(A\act{e}B,\emptyset)
\end{array}}
{R;\Gamma \Gives \s{fix}_{r} f.M: (A\act{e}B,\emptyset)}
\end{equation}
\end{proposition}

For a concrete example, assume basic operators on the integer type and let $M$  be 
the factorial function:
\[
M= \lambda x.\s{if} \ x=0 \ \s{then} \ 1 \ \s{else} \ x*f(x-1)~.
\]
Then compute $(\s{fix}_{r} f.M)1$. In this case
we have $e=\set{r}$ and $r:\w{int} \act{r} \w{int} \in R$.

\section{Types and effects: stratified case}\label{stratified}
As we have seen, an unstratified simply typed calculus with effects
may produce diverging computations. To avoid this, a natural idea 
proposed by G. Boudol in \cite{Boudol07} is to {\em stratify} regions.

Intuitively, we fix a well-founded order on regions and we make sure
that values stored in a region $r$ can only produce effects on smaller
regions.  For instance, suppose $V$ is a value with type
$(\Ter \act{\set{r}} \Ter)$.  Intuitively, this means that when applied to an
argument $U:\Ter$, $V$ may produce an effect on region $\set{r}$. Then
the value $V$ can only be stored in regions larger than $r$. We shall
see that this stratification allows for an inductive definition of the
values that can be stored in a given region.

The only change in the type system concerns the judgements
$R\Gives$, $R\Gives A$, and $R\Gives (A,e)$ whose rules are
redefined as follows:
\[
\begin{array}{c}

\infer{}{\emptyset \Gives}

\qquad

\infer{R\Gives A\quad r\notin \w{dom}(R)}
{R,r:A\Gives} 
\qquad

\infer{R\Gives}
{R\Gives \Ter} \\ \\

\infer{R\Gives \quad r:A \in R}
{R\Gives \regtype{r}{A}} 

\qquad

\infer{R\Gives A\quad R\Gives B\quad e\subseteq \w{dom}(R)}
{R\Gives A\act{e} B} 

\qquad

\infer{R\Gives A\quad e\subseteq\w{dom}(R)}
{R\Gives (A,e)}~.

\end{array}
\]

\paragraph{Proviso}
Henceforth we shall use $\Gives$ to refer to provability in the stratified system
and $\Gives^u$ for provability in the {\em unstratified} one. The former
implies the latter since $R\Gives$ implies $R\Gives^u$ and 
$R\Gives A$ implies $R\Gives^u A$, while the other rules are unchanged.

\subsection{Basic properties revisited}
The main properties we have proven for the unstratified system can 
be specialised to the stratified one.

\begin{proposition}[basic properties, stratified] \label{strat-prop}
The following properties hold in the 
stratified system.

\begin{description}

\item[weakening]
If $R;\Gamma \Gives M:(A,e)$ and $R,R' \Gives \Gamma,\Gamma'$ then
 $R,R';\Gamma,\Gamma' \Gives M:(A,e)$.

\item[substitution]
If $R;\Gamma,x:A \Gives M:(B,e)$ and 
$R;\Gamma \Gives N:(A,\emptyset)$ then 
$R;\Gamma \Gives [N/x]M:(B,e)$.

\item[subject reduction]
If $R,R'; \Gives \vc{M},S:(\Beh,e)$, 
$R \Gives e$, and $\vc{M},S \arrow \vc{M'},S'$  then
$R,R'; \Gives \vc{M'},S':(\Beh,e)$,
$S_{\mid \w{dom}(R')} = S'_{\mid \w{dom}(R')}$, and
$\vc{M},S_{\mid \w{dom}(R)} \arrow \vc{M'},S'_{\mid\w{dom}(R)}$.
Moreover, if $\vc{M}=M$ and $R,R';\Gives M:(A,e)$ then 
$\vc{M'}=M'$ and  $R,R';\Gives M':(A,e)$.
\end{description}
\end{proposition}

\subsection{Interpretation}\label{reducibility}
We describe a {\em reducibility candidates} interpretation that entails that
typed programs terminate.
We denote with $\w{SN}$ the collection of strongly normalising
single-threaded programs, {\em i.e.}, the programs of the shape $M,S$ 
such that all reduction sequences terminate.  
We write $(M,S)\Downarrow(N,S')$ if
$M,S \act{*} N,S'$ and $N,S'\not\arrow$.
We write $R'\geq R$, and say that $R'$ extends $R$,
if $R'\Gives$ and $R'= R,R''$ for some $R''$.

The starting idea is that the interpretation of $R\Gives$ is a set of
stores and the interpretation of $R\Gives (A,e)$ is a set of terms.
One difficulty is that the stores and the terms may depend on a region
context $R'$ which extends $R$.  We get around this problem, by
making the context $R'$ explicit in the interpretation.  Then the
interpretation can be given directly by induction on the provability
of the judgements $R\Gives$ and $R\Gives (A,e)$. 
This is a notable simplification with respect to the 
approach taken in \cite{Boudol07} where 
a rather {\em ad hoc} well-founded order on judgements
is introduced to define the interpretation.

A second characteristic of our approach is that the properties a thread
must satisfy are specified with respect to a `saturated' store which
intuitively already contains all the values the thread may write into
it.  This approach simplifies the interpretation and 
provides a simple argument to extend the
termination argument from single-threaded to multi-threaded programs.
Indeed, if we a have a set of threads which are guaranteed to
terminate with respect to a saturated store then their parallel
composition will terminate too. To see this, one can reason by
contradiction: if the parallel composition diverges then one thread
must run infinitely often and, since the threads cannot modify the
saturated store (what they write is already there), this contradicts
the hypothesis that all the threads taken alone with the saturated
store terminate.

Finally, minor technical differences with respect to \cite{Boudol07}
is that we interpret the subtyping rule (cf. discussion in
section \ref{unstratified}) and that our notion of reducibility candidate 
follows  Girard rather than Stenlund-Tait (see \cite{Gal90} for a detailed comparison
and references).

\begin{description}

\item[Region-context]
Let $R=r_1:A_1,\ldots,r_n:A_n$  and $R_{r_{i}} = r_1:A_1,\ldots,r_{i-1}:A_{i-1}$,
for $i=1,\ldots,n$. 
We interpret a region-context $R$ as a set of pairs $R'\Gives S$ where
$R'$ is a region-context which extends $R$ and $S$ is a `saturated' store
whose domain coincides with $R$:
\[
\begin{array}{lll}
\sem{R} 
&=
&\{ \ R'\Gives S \mid R'\geq R, \quad \w{dom}(S)=\w{dom}(R), \mbox{
  and for } i=1,\ldots,n \\
&&\quad \qquad \qquad  S(r_{i})= \set{V \mid R'\Gives V \in \sem{R_{r_{i}} \Gives
    (A_i,\emptyset)}} 
 \ \}
\end{array}
\]
If $R'\geq R$ then $\sem{R}(R')$ is defined as the store $S$ such 
that $R'\Gives S\in \sem{R}$. Note that, for $r\in \w{dom}(R)$
and $R=R_1,r:A,R_2$, $V\in \sem{R}(R')(r)$ means $R'\Gives V\in 
\sem{R_{1}\Gives (A,\emptyset)}$.

\item[Type and effect]
We interpret a type and effect $R\Gives(A,e)$
as the set of pairs $R'\Gives M$ such that
$R'$ extends $R$, and $M$ is a closed term typable with respect to 
$R'$ and satisfying suitable properties (1-3 below):
\[
\begin{array}{lll}
\sem{R\Gives (A,e)}=
& \{ R'\Gives M \mid 
&(1)\quad R'\geq R, \quad R';\emptyset\Gives M:(A,e),  \\
&&(2)\quad \mbox{for all }R''\geq R' \ 
            M,\sem{R}(R'')\in \w{SN}, \mbox{ and }\\
&&(3)\quad \mbox{for all }M',S',R''\geq R' \quad 
           (M,\sem{R}(R''))\Conv (M',S')  \\
&&\quad\quad\quad\mbox{implies } S'=\sem{R}(R'') \mbox{ and }\cl{C}(A,R,R'',M')  \  \} 
\end{array}
\]
\[
\begin{array}{ll}
\mbox{where: } 
&\cl{C}(A,R,R'',M') \equiv \\ 
&(A=\Ter \quad \impl  \quad M'= *) \ \AND \\
&(A=\regtype{r}{B} \quad \impl \quad  M'=r) \ \AND \\
&(A= A_1\act{e'}A_2 \quad \impl \quad M'=\lambda x.N \quad \AND \\
&\qquad\qquad \qquad\qquad\qquad 
\mbox{for all } \ R_1\geq R'', R_1\Gives V\in \sem{R\Gives (A_1,\emptyset)} \\ 
&\qquad\qquad\quad \qquad\qquad \qquad  \mbox{implies }R_1 \Gives M'V \in \sem{R \Gives (A_2,e')} \  )~.

\end{array}
\]
\end{description}

Suppose $R=r_1:A_1,\ldots,r_n:A_n$. 
We note that the interpretation of
$R$ depends on the interpretation of
$r_1:A_1,\ldots,r_{i-1}:A_{i-1}\Gives A_i$ for $i=1,\ldots,n$ and
the interpretation of $R\Gives (A,e)$ depends on the interpretation
of $R$ and, when $A=A_1\act{e'}A_2$, on the interpretation of
$R\Gives (A_1,\emptyset)$ and $R\Gives (A_2,e')$. 
It is easily verified that the definition of the interpretation
is well founded by considering as measure the height of the proof
of the interpreted judgement. 
We also note that such a well-founded definition would not be possible
in the unstratified system. For instance, the interpretation of 
$r:A\Gives (A,\emptyset)$ where $A=\Ter \act{r} \Ter$ should refer to a store
containing values of type $A$.
Finally, we stress that the interpretations of $R$ and $R\Gives (A,e)$
actually contain terms typable in an extension $R'$ of $R$ but that
their properties are stated with respect to a store whose domain 
is $\w{dom}(R)$. This is possible because the type and effect
system does indeed delimit the effects a term may have when it is 
executed (cf. remark \ref{effect-delimitation-rmk}).

\subsection{Basic properties of the interpretation}
We say that a term $M$ is {\em neutral} if it is not a $\lambda$-abstraction.
The following proposition lists some basic properties of the interpretation.
Similar properties arise in the reducibility candidates
interpretations used for `pure' functional languages, but the main point
here is that we have to state them relatively to suitable stores.
In particular, the extension/restriction property, which is perhaps
less familiar, is crucial to prove the following soundness theorem 
\ref{main-thm}.

\begin{proposition}[properties interpretation]\label{int-prop}
The following properties hold.

\begin{description}

\item[Weakening] If $R'' \geq R'\geq R$, $R\Gives (A,e)$, and 
$R'\Gives M\in \sem{R \Gives (A,e)}$ then 
$R''\Gives M\in \sem{R \Gives (A,e)}$.

\item[Extension/Restriction] Suppose $R'' \geq R'\geq R$ and $R\Gives (A,e)$.
Then  $R''\Gives M\in \sem{R \Gives (A,e)}$ if and only if 
$R''\Gives M\in \sem{R' \Gives (A,e)}$.

\item[Subtyping] 
If $R\Gives (A,e)\leq (A',e')$ then 
$\sem{R\Gives (A,e)} \subseteq \sem{R\Gives (A',e')}$.

\item[Strong normalisation] If $R'\Gives M \in \sem{R\Gives (A,e)}$  and 
$R''\geq R'$ then $M,\sem{R}(R'') \in \w{SN}$.

\item[Reduction closure]
If $R'\Gives M \in \sem{R\Gives (A,e)}$, $R''\geq R'$,
and $M,\sem{R}(R'') \arrow M',S'$ then 
$R''\Gives M'\in \sem{R\Gives (A,e)}$ and $S'=\sem{R}(R'')$.

\item[Non-emptiness] 
If $R\Gives A$ then there is a value  $V$ such 
that for all $R'\geq R$ and $e\subseteq \w{dom}(R)$, 
$R'\Gives V\in \sem{R\Gives (A,e)}$.

\item[Expansion closure] 
Suppose $R\Gives (A,e)$, $R'\geq R$, $R';\emptyset \Gives M:(A,e)$, and $M$ is
neutral. Then $R'\Gives M\in \sem{R\Gives (A,e)}$ provided that
for all $R''\geq R',M',S'$ such that $M,\sem{R}(R'')\arrow M',S'$ we have
that $R''\Gives M'\in \sem{R\Gives (A,e)}$ and $S'= \sem{R}(R'')$.

\end{description}

\end{proposition}
\Proofhint
\begin{description}

\item[Weakening] 
We rely on proposition \ref{strat-prop}((syntactic) weakening) and
the fact that, the properties the pairs $R'\Gives M$ must satisfy to belong to
$\sem{R\Gives (A,e)}$, must hold for all the extensions $R''\geq R'$.

\item[Extension/Restriction] 
By definition,  $\sem{R}(R'')$ coincides
with $\sem{R'}(R'')$ on $\w{dom}(R)$. On the other hand, the
proposition \ref{strat-prop}(subject reduction) guarantees that the
reduction of a term of type and effect $(A,e)$ will not depend and
will not affect the part of the store whose domain is
$\w{dom}(R')\minus\w{dom}(R)$.
We then prove the property by induction on the structure of the type $A$.

\item[Subtyping] 
This is proven by induction on the the proof of $R\Gives A\leq A'$.

\item[Strong normalisation] 
This follows immediately from the definition of the interpretation.

\item[Reduction closure]
We know that $M,\sem{R}(R'')$ must normalise to a value 
satisfying suitable properties  and the
same saturated store $\sem{R}(R'')$. 
Moreover, we know that the store can only grow during 
the reduction. We conclude applying the weakening property.

\item[Non-emptiness/Expansion closure] 
These two properties are proven at once, by induction
on the proof height of $R\Gives (A,e)$. We take as values:
$*$ for the type $\Ter$, $r$ for a type of the shape $\regtype{r}{B}$,
and the `constant function' $\lambda x.V_2$ for a type of the shape 
$A_{1} \act{e_{1}} A_2$ where $V_2$ is the value inductively built for
$A_2$. To prove $\lambda x.V_2 \in 
\sem{R\Gives (A_1\act{e_{1}} A_2,e)}$, we use the inductive
hypothesis of expansion closure of $\sem{R\Gives (A_2,e_{1})}$. \qed
\end{description}

\subsection{Soundness of the interpretation}
By definition, if $R\Gives M \in \sem{R\Gives (A,e)}$ then 
$R; \Gives M:(A,e)$. We are going to show that the converse
holds too. First we need to generalise the notion of
reducibility to open terms.

\begin{definition}[term interpretation]
We write  $R;x_1:A_1,\ldots,x_n:A_n \Models M:(B,e)$ if 
whenever $R'\geq R$ and 
$R'\Gives V_i \in \sem{R \Gives (A_i,\emptyset) }$ for  $i=1,\ldots,n$ 
we have that  $R'\Gives [V_1/x_1,\ldots,V_n/x_n]M \in \sem{R \Gives (B,e)}$.
\end{definition}

As usual, the main result can be stated as the soundness of the
interpretation with respect to the typing rules. Since terms in the
interpretation are strongly normalising relatively to a saturated
store (cf. proposition \ref{int-prop}),  it follows that 
typable (closed) terms are strongly normalising.

\begin{theorem}[soundness]\label{main-thm}
If  $R;\Gamma \Gives M:(B,e)$ then 
$R;\Gamma \Models M:(B,e)$.
\end{theorem}
\Proofhint
The proof goes by induction on the typing of the terms
and exploits the properties of the interpretation stated in 
proposition \ref{int-prop}. As usual, 
the case of the abstraction is proven by
appealing to expansion closure and the case of application
follows from the very interpretation of the functional
types and reduction closure. The cases where we write or read from the store
have to be handled with some care. 
We discuss a simplified situation. Suppose $R'\geq R=R_1,r:A,R_2$.

\begin{description}

\item[write] Suppose $R;\Gives \st{r}{V}:(\Ter,\set{r})$ 
is derived from $R;\Gives V:(A,\emptyset)$. 
Then, by induction hypothesis, we know that 
$R'\Gives V \in \sem{R\Gives (A,\emptyset)}$.
However, for maintaining the invariant that the
saturated store is unchanged, we need to show that
$R'\Gives V \in \sem{R_1 \Gives (A,\emptyset)}$,
and this is indeed the case thanks to 
proposition \ref{int-prop}(restriction).

\item[read]
Suppose we have $R';\Gives \gt{r}:(A,\set{r})$. 
Now notice that proposition \ref{int-prop}(non-emptiness)
guarantees that $\sem{R}(R')(r)$ is not empty. 
Thus $\gt{r},\sem{R}(R')$ will reduce to 
$V,\sem{R}(R')$ for some value $V$ such that
$R'\Gives V \in \sem{R_{1}\Gives (A,\emptyset)}$.
However, what we need to show is that
$R'\Gives V \in \sem{R \Gives (A,\emptyset)}$ 
and this is indeed the case thanks to 
proposition \ref{int-prop}(extension). \qed
\end{description}

\begin{corollary}[termination]\label{cor-sound}
\Defitemf{(1)}
The judgement $R; \Gives M:(A,e)$ is provable if and only if
$R\Gives M\in \sem{R\Gives (A,e)}$.

\Defitem{(2)} Every typable multi-threaded program 
$R; \Gives M_1,\ldots,M_n: (\Beh,e)$ 
terminates.
\end{corollary}

Corollary \ref{cor-sound}(1), follows from theorem \ref{main-thm} taking the context
$\Gamma$ to be empty. Corollary \ref{cor-sound}(2) follows from the fact that each
thread strongly normalizes with respect to a saturated store. Then its
execution is not affected by the execution of other threads in
parallel: all these parallel threads could do is to write in the
saturated store values which are already there.

\section{Extensions}\label{extension-sec}
In this section we sketch two extensions of our
basic model.  
The first simple one (section
\ref{thread-sec}) concerns the possibility of generating 
dynamically new threads while the second (section \ref{time-sec})
is a bit more involved and it concerns the notion of timed/synchronous
computation.

\subsection{Thread generation}\label{thread-sec}
In the presented system, the number of threads is constant. We
describe a simple extension that allows to generate new
threads during the execution.  Namely, (1) we  regard a
multi-set of terms $M_1,\ldots,M_n$ as a term of behaviour type $\Beh$ and 
(2) we abstract terms of behaviour type $\Beh$ producing 
terms of type $(A\act{e} \Beh)$ for some type $A,e$ 
(this formalisation is inspired by \cite{AC98}(chpt. 16)). 
It is straightforward to extend the rules for the formation
of region contexts and types and for subtyping to take into account
the behaviour type $\Beh$.  Similarly, the typing rules for
abstraction and application are extended to take into account the
situation where the codomain of the functional space is $\Beh$. 
The full definition of this system is given in appendix \ref{summary-sec}.
In this extended system, we can then type, {\em e.g.}, 
a term that after performing
an input will start two threads in parallel: 
$(\lambda x.(M,N))\gt{r}$ which would be written in, say, the
$\pi$-calculus as $r(x).(M\mid N)$.

In order to show termination of this extended language, we have
to define the interpretation of the judgement $R\Gives (\Beh,e)$.
To this end, it is enough to extend the definition in section 
\ref{reducibility} by requiring that  a term in 
$\sem{R\Gives (\Beh,e)}$ when run
in the saturated store will indeed terminate without 
modifying the store and produce a multi-set of values.
Formally, we add the condition `$A = \Beh \impl M'= V_1,\ldots,V_n, n\geq 1$'
to the definition of the predicate $\cl{C}$.
We can then lift our results to this system leaving the 
structure of the proofs unchanged.

\subsection{Synchrony/Time}\label{time-sec}
We consider a timed/synchronous extension of our language.  
Following an established tradition, we
consider that the computation is divided into {\em instants} and that
an instant ends when the computation cannot progress. Then we need at
least an additional operator that allows to write programs that {\em
react} to the end of the instant by changing their state in the
following instant. 
We shall see that the termination of the typable programs can be obtained
by mapping reductions in the extended language into reductions in the
core language.

\paragraph{Syntax and Reduction}
We extend the collection of terms as follows:
$M::= \cdots \Alt \pres{M}{M}$, where
the operator \w{else-next}, written $\pres{M}{N}$, tries to run
$M$ and, if it fails, runs $N$ in the following instant (cf. \cite{HR95}).
We extend the evaluation contexts assuming:
$E::= \cdots \Alt \pres{E}{M}$,
and the {\em elementary} evaluation contexts assuming:
$\w{El}::= \cdots \Alt \pres{[~]}{M}$.

We define a simplification operator \w{red} that removes from a context all
pending branches else-next:
\[
\w{red}(E) = \left\{
\begin{array}{ll}
[~] &\mbox{if }E=[~] \\
\w{red}(E') &\mbox{if }E=\pres{E'}{N}  \\
\w{El}[\w{red}(E')]  &\mbox{otherwise, if } E=\w{El}[E'] 
\end{array}\right.
\]
We say that an evaluation context $E$ is time insensitive if 
$\w{red}(E)=E$.
We adapt the reduction rules defined in section \ref{lambda-region}
as follows:
\[
\begin{array}{lll}

E[(\lambda x.M)V] &\arrow &\w{red}(E)[[V/x]M] \\
E[\gt{r}],\store{r}{V} &\arrow &\w{red}(E)[V],\store{r}{V} \\ 
E[\st{r}{V}]           &\arrow &\w{red}(E)[*],\store{r}{V}~.
\end{array}
\]
Further, we have to describe how a program reacts to the end of
the computation. This is specified by the 
relation $\act{\tick}$ below:
\[
\begin{array}{c}

\infer{}
{V\act{\tick} V}
\qquad

\infer{}
{S\act{\tick} S}
\qquad
\infer{M=E[\gt{r}]\quad E \mbox{ time insensitive}}
{M \act{\tick} M}
\\ \\

\infer{\begin{array}{c}
M=E[\pres{E'[\Delta]}{N}] \quad
\Delta::=V\Alt \gt{r} \\
E \mbox{ time insensitive}
\end{array}}
{M \act{\tick} E[N]} 

\qquad

\infer{\begin{array}{c}
P_i \act{\tick} P'_i\quad i=1,2 \\
P_1,P_2\not\arrow
\end{array}}
{P_1,P_2 \act{\tick} P'_{1},P'_{2}}~.

\end{array}
\]
For instance, we can write $\pres{(\lambda x.M)\gt{r}}{N}$ for a thread that
tries to read a value from the region $r$ in the first instant and if it fails
it resumes the computation with $N$ in the following instant.
We can also write $\pres{*}{N}$ for a thread that (unconditionally) stops its computation for
the current instant and resumes it with $N$ in the following instant.

Note that $P\act{\tick}$ only if $P\not\arrow$.  The converse is in
general false, but it holds for well-typed closed programs
(cf. proposition \ref{decomp-revisited}). Thus for well-typed closed programs
the principle is that time passes (a $\act{\tick}$ transition is
possible) exactly when the computation cannot progress (a $\arrow$
transition is impossible). Then termination is obviously a {\em very}
desirable property of timed/synchronous programs.

\paragraph{Typing}
The typing rules for the terms are extended as follows:
\[
\begin{array}{c}

\infer{R;\Gamma\Gives M:(A,e)\quad R;\Gamma \Gives N:(A,e')}
{R;\Gamma \Gives \pres{M}{N} : (A,e)}~.

\end{array}
\]
Note that  
in typing $\pres{M}{N}$ we only record the
effect of the term $M$, that is we focus on the effects a term
may produce in the first instant while neglecting those that
may be produced at later instants.

\paragraph{Reduction}
The decomposition proposition \ref{decomp} can be lifted 
to the extended language. There is a third case to be considered
besides the two  arising in proposition \ref{decomp}
which corresponds to the situation where the redex is under the
scope of an else-next. More precisely, in the third case
a closed term $M$ is decomposed as 
$E[\pres{E'[\Delta]}{N}]$ where $E$ is a time insensitive
evaluation context and $\Delta$ has the shape $V$, 
$(\lambda x.N)V$, $\st{r}{V}$, or $\gt{r}$. 

Focusing on the stratified case, one can adapt 
the weakening, substitution, and subject reduction
properties whose proofs proceed as in proposition  \ref{strat-prop}.
The preservation of the type information by the passage 
of time (tick reduction) can be stated as follows.
\begin{quote}
If $R; \Gives \vc{M},S:(\Beh,e)$, 
and $\vc{M},S \act{\tick} \vc{M'},S'$  then
$S=S'$ and there is an effect $e'$ such that
$R; \Gives \vc{M'},S:(\Beh,e')$.
\end{quote}
Notice that the effect of the reduced term might be incomparable 
with the effect of the
term to be reduced. 
Still the following {\em context substitution} property 
allows to conclude that the resulting term is well-typed.
\begin{quote}
If $R;\Gamma,x:A \Gives E[x]:(B,e)$ where $x$ is not free in the evaluation context $E$ and 
$R;\Gamma \Gives N:(A,e')$ then 
$R;\Gamma \Gives E[N]:(B,e\union e')$.
\end{quote}

\paragraph{Translation}
We consider a translation
that removes the else-next operator while preserving
typing and reduction. Namely, we define a function 
$\tra{\_}$ on terms such that 
$\tra{\pres{M}{N}}=\tra{M}$, $\tra{x}=x$, $\tra{*}=*$,
$\tra{r}=r$, and  which commutes with the other operators (abstraction, application,
reading, and writing).
Also the translation is extended to stores and programs in the obvious way:
$\tra{\store{r}{V}} = \store{r}{\tra{V}}$,
$\tra{X_1,\ldots,X_n} = \tra{X_{1}},\ldots,\tra{X_{n}}$.

\begin{proposition}[simulation]\label{else-next-elim}
\Defitemf{(1)}
If $R;\Gamma \Gives M:(A,e)$ then 
$R;\Gamma \Gives \tra{M}:(A,e)$.

\Defitem{(2)}
If $R;\Gamma \Gives P:(\Beh,e)$ then 
$R;\Gamma \Gives \tra{P}:(\Beh,e)$.

\Defitem{(3)}
If $R; \Gives P:(\Beh,e)$ and $P\arrow P'$
then $\tra{P}\arrow \tra{P'}$.

\Defitem{(4)} 
A program $P$ terminates if $\tra{P}$ terminates.

\end{proposition}

The proof of this proposition is direct. In particular, to prove (3) we
show that the translation commutes with the substitution
and that the translation of an evaluation context is again an
evaluation context.

\paragraph{Fixed-point, revisited}
The typing rule (\ref{fix-type-unstratified}) proposed for the
fixed-point combinator cannot be applied in the stratified system as
the condition $r:A\act{e}B\in R$ and $r\in e$ cannot be satisfied.
However, we can still type recursive calls that happen in a later
instant. 

\begin{proposition}[type fixed-point, revisited]\label{derived-rule2-fix}
The following typing rule for the fixed point combinator is derived in
the stratified system
\begin{equation}\label{fix-type-stratified}
\infer{\begin{array}{c}
R;\Gamma,f:A\act{e\union \set{r}}B\Gives  M:(A\act{e}B,\emptyset)
\qquad r:A\act{e}B \in R
\end{array}}
{R;\Gamma \Gives \s{fix}_{r} f.M: (A\act{e\union \set{r}}B,\emptyset)}
\end{equation}
\end{proposition}

We prove this proposition by a direct application of 
the typing rules and the substitution property 
(cf. proposition \ref{basic-prop-extended}).
To see a concrete example where the rule can be applied, consider a
thread that at each instant writes an integer in a region $r'$ (we assume
a basic type \w{int} of integers):
\[
M= \lambda x.(\lambda z.\pres{*}{f(x+1)})(\st{r'}{x})
\] 
Then, {\em e.g.}, $(\s{fix}_r f.M)1$
is the infinite behaviour that at the $i$-${\w{th}}$ instant writes $i$ in region $r'$.
One can check the typability of $\s{fix}_r f.M$ taking as (stratified) 
region context
$R=r':\w{int},r:\w{int}\act{\set{r'}} \Ter$.

\section{Conclusion}
We have introduced a $\lambda$-calculus with regions as an 
abstraction of a variety of concrete higher-order concurrent languages
with specific scheduling and interaction mechanisms.
We have described a stratified type and effect system and provided
a new reducibility candidates interpretation for it which entails that
typable programs terminate.

We have highlighted some relevant properties of the interpretation
(proposition \ref{int-prop}) which could be taken as the basis for an
abstract definition of reducibility candidate. The latter is needed to
interpret second-order (polymorphic) types (see, {\em e.g.},
\cite{Gal90}).  We believe the proposed proof is both more general
because it applies to a variety of interaction mechanisms and
scheduling policies and simpler to understand because the
interpretation is given by a direct induction on the proof system and
because the invariant on the store is easier to manage (the store 
is not affected by the reduction).  
This is of course a subjective opinion and the reader
who masters \cite{Boudol07} may well find our revised treatment
superfluous.

We have also lifted our approach to a timed/synchronous framework
and derived a form of recursive definition which is useful to define
behaviours spanning infinitely many instants.

In ongoing work, we have refined the type and effect system to 
include {\em linear information} (in the sense of linear logic) 
which is relevant both to define deterministic fragments of the calculus 
and to control better the complexity of the definable programs.

{\footnotesize
\paragraph{Acknowledgements}
Thanks to G\'erard Boudol for several discussions on \cite{Boudol07}.

}

\newpage

\appendix

\section{Proofs}\label{proofs-sec}

\subsection{Proof of proposition \ref{decomp} (decomposition)}\label{proof-decomp}
By induction on the structure of $M$.  
By the typing hypothesis, $M$ cannot be a variable.
If $M$ is a value we are in case 1.
Otherwise, $M$ can have exactly one of the following shapes:
$M_1M_2$, $\gt{M_{1}}$, $\st{M_{1}}{M_{2}}$.
We consider in some detail the case for application.

The typing rules force $M_1$ and $M_2$ to be typable 
in an empty context. Moreover $M_1$ must have a functional type.
Because of this, if $M_1$ is a value then it must be of the shape $\lambda x.M'_1$.
Moreover, we can apply the inductive hypothesis to $M_2$ and suitably compose
with the evaluation context $M_1[~]$.
If $M_1$ is not a value then we apply the inductive hypothesis to $M_1$ and
suitably compose with the evaluation context $[~]M_2$. \qed

\subsection{Proof of proposition \ref{basic-prop-unstratified} (basic
  properties, unstratified)}\label{proof-basic-prop}
\begin{description}

\item[Weakening]
First prove by induction on the proof height 
that if  $R,R'\Gives$ and 
$R\Gives A$, ($R\Gives (A,e)$, $R\Gives A\leq B$) then
$R,R'\Gives A$ ($R,R'\Gives (A,e)$, $R,R'\Gives A\leq B)$.
Next, by induction on the proof height, we show how to transform a proof 
$R;\Gamma \Gives M:(A,e)$ into a proof of 
$R,R';\Gamma,\Gamma'\Gives M:(A,e)$. \qed

\item[Substitution]
By induction on the proof height  of 
$R;\Gamma,x:A \Gives M:(B,e)$. 

\item[Subject reduction]
First we notice that if a term $M$, of type and effect $(A,e)$, is
ready to interact with the store then the region on which
the interaction takes place belongs to $e$. More formally, if 
$R;\Gives M:(A,e)$, $M\equiv E[\Delta]$ and $\Delta$ has
the shape $\gt{r}$ or $\st{r}{V}$ then $r\in e$.
To prove these facts we proceed by induction on the structure of the
evaluation context $E$.
Then we prove the assertion by case analysis on the 
reduction rule applied relying on the substitution property. \qed 

\end{description}

\subsection{Proof of proposition \ref{derived-rule1-fix} (type fixed-point)}
Suppose $r:A\act{e}B \in R$ and $r\in e$. 
Then $R; \Gives \lambda x.\gt{r}x: (A\act{e}B,\emptyset)$.
By  proposition \ref{basic-prop-unstratified}(substitution),
$R;\Gamma\Gives M': (A\act{e}B,\emptyset)$ where
$M'=[\lambda x.\gt{r}x/f]M$.
From this we derive:
$R;\Gamma\Gives M'':(A\act{e}B,\set{r})$ where
$M''=\gt{\regterm{r}{\lambda x.M'x}}$. 
This judgement can be weakened to $R;\Gamma,x:A\Gives M'':(A\act{e}B,\set{r})$
which combined with $R;\Gamma,x:A\Gives x:(A,\emptyset)$ leads to
$R;\Gamma\Gives \lambda x.M''x:(A\act{e}B,\emptyset)$ where
$\lambda x.M''x= \s{fix}_{r} f.M$, as required. \qed

\subsection{Proof of proposition \ref{int-prop} (properties interpretation)}

\begin{description}

\item[Weakening] 
Suppose $R''\geq R'\geq R$ and $R'\Gives M\in \sem{R\Gives (A,e)}$.
Then $R';\emptyset \Gives M:(A,e)$ and by proposition
\ref{strat-prop}(weakening) we know that $R'';\emptyset\Gives
M:(A,e)$.  Moreover, an inspection of the definition of $\sem{R\Gives
(A,e)}$ reveals that if we take a $R'''\geq R''$ then the required
properties are automatically satisfied because $R'''\geq R'$ and
$R'\Gives M\in \sem{R\Gives (A,e)}$.

\item[Extension/Restriction] 
Suppose $R''\geq R'\geq R$ and $R\Gives (A,e)$.
We want to show that:
\[
R''\Gives M\in \sem{R\Gives (A,e)}
\mbox{ iff }
R''\Gives M\in \sem{R'\Gives (A,e)}~.
\]
Note that $\sem{R'}(R'')$ coincides with $\sem{R}(R'')$ on
$\w{dom}(R)$. On the other hand, the 
proposition \ref{strat-prop}(subject reduction) guarantees
that the reduction of a term of type and effect $(A,e)$
will not depend and will not affect the part of the
store whose domain is $\w{dom}(R')\minus\w{dom}(R)$.

We proceed by induction on the structure of the type $A$.

Suppose $A=\Ter$.
If $R''\Gives M\in \sem{R\Gives (A,e)}$ then we know
that for any $R_1\geq R''$ we have that 
$M,\sem{R}(R_1)$ strongly normalizes to $*,\sem{R}(R_1)$.
By applying subject reduction, we can conclude
that $M,\sem{R'}(R_1)$ will also strongly normalize
to $*,\sem{R'}(R_{1})$.
A similar argument applies if 
we start with  $R''\Gives M\in \sem{R'\Gives (A,e)}$.
Also, this proof schema can be repeated if $A=\regtype{r}{B}$.

Suppose now $A=A_1\act{e_{1}} A_2$. 
If $R''\Gives M\in \sem{R\Gives (A,e)}$ 
then we know that for any $R_1\geq R''$,
$M,\sem{R}(R_1)$ strongly normalizes to $\lambda x.N,\sem{R}(R_1)$,
for some $\lambda x.N$. 
Moreover for any $R_2\geq R_1$, we have that
$R_2\Gives V\in \sem{R\Gives (A_1,\emptyset)}$
implies $R_2\Gives (\lambda x.N)V\in \sem{R\Gives (A_2,e_1)}$.
By applying subject reduction, we can conclude
that $M,\sem{R'}(R_1)$ will also strongly normalize
to $(\lambda x.N),\sem{R'}(R_{1})$, for some
value $\lambda x.N$.
Further, by induction hypothesis on $A$,
if $R_2\geq R_1$ and 
$R_2\Gives V \in \sem{R'\Gives (A_1,\emptyset)}$
then  $R_2\Gives (\lambda x.N)V\in \sem{R'\Gives (A_2,e_1)}$.

Again, a similar argument applies if 
we start with  $R''\Gives M\in \sem{R'\Gives (A,e)}$.

\item[Subtyping] 
Suppose $R\Gives (A,e)\leq (A',e')$. We proceed by induction
on the proof of $R\Gives A\leq A'$. 

Suppose we use the axiom $R\Gives A\leq A$ and
$R'\Gives M\in \sem{R\Gives (A,e)}$. 
Then we check that $R'\Gives M\in \sem{R\Gives (A,e')}$ since
$R';\emptyset\Gives M:(A,e')$ using the subtyping rule,
and the remaining conditions do not depend on $e$ or $e'$.

Suppose we have $A=A_1\act{e_{1}} A_2$, $A'= A'_1\act{e'_{1}} A'_2$,
and we derive  $R\Gives A\leq A'$ from
$R\Gives A'_1\leq A_1$, $R\Gives A_2\leq A'_2$, and $e_1\subseteq e'_1$.
Moreover, suppose $R'\Gives M\in \sem{R\Gives (A,e)}$.
Then
$R';\emptyset\Gives M:(A',e')$, by the subtyping rule.
Moreover, if $R''\geq R'$ and $M,\sem{R}(R')$ reduces to 
$\lambda x.N,\sem{R}(R'')$, 
we can use the induction hypothesis to show that
if $R_1\geq R''$ and $R_1\Gives V\in \sem{R\Gives (A'_{1},\emptyset)}$
then $R_1\Gives (\lambda x.N)V\in \sem{R\Gives (A'_{2},e'_{1})}$.

\item[Strong normalisation] 
This follows immediately from the definition of $\sem{R\Gives (A,e)}$.

\item[Reduction closure]
Suppose $R'\Gives M\in \sem{R\Gives (A,e)}$ and $R''\geq R'$.
We know that $M,\sem{R}(R'')$ strongly normalizes to programs of the shape
$M'',\sem{R}(R'')$ where $M''$ has suitable properties. 
Then if $M,\sem{R}(R'')$ reduces to $M',S'$ 
it must be that $S'=\sem{R}(R'')$
since the store can only grow. 
Moreover, by proposition \ref{strat-prop}(subject reduction),
we know that $R'';\emptyset \Gives M':(A,e)$. 
It remains to check conditions (2) and (3) of the interpretation 
on $R''\Gives M'$.
Let $R'''\geq R''$. We claim $M,\sem{R}(R''')$ reduces to 
$M',\sem{R}(R''')$ so that $M'$ inherits from $M$ the conditions 
(2) and (3). To check the claim, recall that
$M,\sem{R}(R'')$ reduces to $M',\sem{R}(R'')$. 
Then we analyse the type of reduction performed. 
The interesting case arises when $M$ reads a value $V$ from the store
$\sem{R}(R'')$ where, say, $R''\Gives V\in \sem{R_1 \Gives
  (B,\emptyset)}$  and $R=R_1,r:B,R_2$.
But then we can apply weakening to conclude that 
$R'''\Gives V\in  \sem{R_1\Gives (B,\emptyset)}$.

\item[Non-emptiness/Expansion closure] 
We prove the two properties at once, by induction on the
proof height of $R\Gives (A,e)$.

\begin{itemize}
\item Suppose $R\Gives (\Ter,e)$.  We take $V=*$. 
Then for $R'\geq R$ we have $R';\emptyset\Gives *:(\Ter,e)$.
Also, for any $R''\geq R'$, $*,\sem{R}(R'')$ converges to itself
and satisfies the required properties.
Therefore $R'\Gives * \in \sem{R\Gives (\Ter,e)}$.

This settles non-emptiness. To check expansion closure, 
suppose $R'\geq R$, $R';\emptyset \Gives M:(\Ter,e)$, and
$R''\geq R'$.
By the decomposition proposition \ref{decomp}, 
$M$ is either a value or a term of the shape $E[\Delta]$ where $\Delta$ is a redex.

If $M,\sem{R}(R'')$ does not reduce then $M$ must be the value $*$.
Indeed, by the typing hypothesis it cannot be a region or an abstraction.
Also, it cannot be of the shape $E[\gt{r}]$.
Indeed, suppose $R=R_1,r:B,R_2$, then by induction
hypothesis on $R_1\Gives (B,\emptyset)$, 
we know that the store $\sem{R}(R'')$ contains at least a value
in the region $r$ .

If $M,\sem{R}(R'')$ does reduce then, by hypothesis, for all $M',S'$ such
that $M,\sem{R}(R'')\arrow M',S'$ we have that 
$R''\Gives M'$ belongs to $\sem{R\Gives (A,e)}$ and $S'= \sem{R}(R'')$.
This is enough to check the conditions (2) and (3) of the
interpretation and conclude
that $R'\Gives M$ belongs to $\sem{R\Gives (A,e)}$.

\item The other basic case is $R\Gives (\regtype{r}{B},e)$.
Then we take as value $V=r$ and we reason as in the previous case.

\item Finally, suppose $R\Gives (A_1\act{e_{1}} A_2,e)$.
By induction hypothesis on $R\Gives (A_2,e_1)$, 
we know that there is a value $V_2$ such that for
any $R'\geq R$ we have $R'\Gives V_2\in \sem{R\Gives  (A_2,e_1)}$.
Then we claim that:
\[
R'\Gives \lambda x.V_2 \in \sem{R\Gives (A_1\act{e_{1}} A_2,e)}~.
\]
First, $R';\Gives \lambda x.V_2:  (A_1\act{e_{1}} A_2,e)$ 
is easily derived from the hypothesis that
$R';\Gives V_2 :(A_2,e_{1})$.
The second property of the interpretation is trivially
fulfilled since $\lambda x.V_2$ cannot reduce.
For the third property, suppose $R_1\geq R''\geq R'$ 
and $R_1\Gives V\in \sem{R\Gives (A_1,\emptyset)}$.
We have to check 
that $R_1\Gives (\lambda x.V_2)V$ belongs to $\sem{R\Gives  (A_2,e_1)}$.
We observe that $R_1; \Gives (\lambda x.V_2)V: (A_2,e_1)$,
and the term $(\lambda x.V_2)V$ is neutral.
Moreover, for $R_2\geq R_1$, 
$(\lambda x.V_2)V,\sem{R}(R_2) \arrow V_2,\sem{R}(R_2)$.
Thus we are in the situation to apply the inductive hypothesis
of expansion closure on $R\Gives (A_2,e_1)$. 

This settles non-emptiness at higher-order. To check expansion
closure, suppose $R'\geq R$, $R';\Gives M:(A_1\act{e_{1}} A_2,e)$, and
$M$ neutral.  Then $M$ cannot be a value and for any $R''\geq R'$ the
program $M,\sem{R}(R'')$ must reduce. Indeed, $M$ cannot be stuck on 
a read because if $r\in \w{dom}(R)$ then we know, by inductive
hypothesis, that $\sem{R}(R'')(r)$ is not-empty.  Then we conclude
that $R'\Gives M$ satisfies properties (2) and (3) of the
interpretation because all the terms it reduces to satisfy them. \qed

\end{itemize}

\end{description}

\subsection{Proof of theorem \ref{main-thm} (soundness)}
We proceed by induction on the proof of
$R;\Gamma\Gives M:(B,e)$.
We shall write $[\vc{V}/\vc{x}]$ for
$[V_1/x_1,\ldots,V_n/x_n]$. 
Suppose  $\Gamma=x_1:A_1,\ldots,x_n:A_n$,
$R\Gives \Gamma$ and $R'\geq R$.
We let $R'\Gives \vc{V} \in \sem{R\Gives \Gamma}$ 
stand for $R'\Gives V_i \in \sem{R\Gives (A_i,\emptyset)}$ for
$i=1,\ldots,n$, where $\vc{V}=V_1,\ldots,V_n$.

\begin{itemize}

\item Suppose $\Gamma= x_1:A_1,\ldots,x_i:A_i,\ldots,x_n:A_n$, 
$R;\Gamma \Gives x_i:(A_i,\emptyset)$, $R'\geq R$, 
and $R'\Gives \vc{V} \in \sem{R\Gives \Gamma}$.
Then $[\vc{V}/\vc{x}]x_i =V_i$ and,
by hypothesis, $R'\Gives V_i \in \sem{R\Gives (A_i,\emptyset)}$.

\item Suppose 
$R;\Gamma \Gives * :(\Ter,\emptyset)$, $R'\geq R$, 
and $R'\Gives \vc{V} \in \sem{R\Gives \Gamma}$.
Then $[\vc{V}/\vc{x}]* =*$ and we know that
$R'\Gives * \in \sem{R\Gives (\Ter,\emptyset)}$.

\item  Suppose 
$R;\Gamma \Gives r:(\regtype{r}{B},\emptyset)$, $R'\geq R$, 
and $R'\Gives \vc{V} \in \sem{R\Gives \Gamma}$.
Then $[\vc{V}/\vc{x}]r =r$ and we know that
$R'\Gives r \in \sem{R\Gives (\regtype{r}{B},\emptyset)}$.

\item Suppose 
$R;\Gamma \Gives M:(A',e')$ is derived from
$R;\Gamma \Gives M:(A,e)$ and $R\Gives (A,e)\leq (A',e')$.
Moreover, suppose $R'\geq R$, 
and $R'\Gives \vc{V} \in \sem{R\Gives \Gamma}$.
By induction hypothesis, 
$R'\Gives [\vc{V}/\vc{x}]M \in \sem{R\Gives (A,e)}$.
By proposition \ref{int-prop}(subtyping), we conclude
that $R'\Gives [\vc{V}/\vc{x}]M \in \sem{R\Gives (A',e')}$.

\item Suppose
$R;\Gamma \Gives \lambda x.M:(A\act{e}B,\emptyset)$ is derived from
$R;\Gamma,x:A \Gives M:(B,e)$.
Moreover, suppose $R'\geq R$, 
and $R'\Gives \vc{V} \in \sem{R\Gives \Gamma}$.
We need to check that 
$R'\Gives \lambda x.[\vc{V}/\vc{x}]M$ belongs to 
$\sem{R\Gives(A\act{e}B,\emptyset)}$. 
Namely, assuming 
$R'' \geq R'\geq R$ and $R'' \Gives V \in \sem{R\Gives (A,\emptyset)}$,
we have to show that 
$R'' \Gives [\vc{V}/\vc{x}](\lambda x.M)V \in \sem{R\Gives (B,e)}$.
We observe that $R'';\Gives  [\vc{V}/\vc{x}](\lambda x.M)V :(B,e)$ and
that, by weakening $R'$ to $R''$ and induction hypothesis, we know that 
$R'' \Gives [\vc{V}/\vc{x},V/x]M \in \sem{R\Gives (B,e)}$.
Then we conclude by applying proposition \ref{int-prop}(expansion closure).

\item Suppose $R;\Gamma \Gives MN:(B,e_1\union e_2\union e_3)$
is derived from  $R;\Gamma \Gives M : (A\act{e_{1}}B,e_{2})$ 
and $R;\Gamma \Gives N:(A,e_3)$.
Moreover, suppose $R'\geq R$, 
and $R'\Gives \vc{V} \in \sem{R\Gives \Gamma}$.
By induction hypothesis, we know that 
$R'\Gives [\vc{V}/\vc{x}]M \in \sem{R\Gives (A\act{e_{1}}B,e_{2})}$ 
and $R'\Gives [\vc{V}/\vc{x}]N \in \sem{R\Gives (A,e_{3})}$.
We have to show that:
$R'\Gives [\vc{V}/\vc{x}](MN) \in 
\sem{R\Gives (B,e_1\union e_2\union e_3)}$.
Suppose $R''\geq R'$. Then 
$[\vc{V}/\vc{x}]M,\sem{R}(R'')$ normalizes
to $\lambda x.M',\sem{R}(R'')$ for some value $\lambda x.M'$ and
$[\vc{V}/\vc{x}]N,\sem{R}(R'')$ normalizes to 
$V,\sem{R}(R'')$ for some value $V$. 
Further, by reduction closure, we know that 
$R''\Gives \lambda x.M'\in \sem{(A\act{e_{1}}B,e_{2})}$ and 
$R''\Gives V \in \sem{(A,e_3)}$. 
It is easily checked that the latter implies 
$R''\Gives V\in \sem{(A,\emptyset)}$.
By condition (3) of the interpretation, we derive that 
$R''\Gives (\lambda x.M')V \in \sem{R\Gives (B,e_{1})}$
which suffices to conclude.

\item Suppose $R;\Gamma \Gives \st{M}{N} :(\Ter,e_1\union e_2\union \set{r})$ is
derived from $R;\Gamma \Gives M:(\regtype{r}{A},e_1)$ and
$R;\Gamma \Gives N:(A,e_2)$.
Moreover, suppose $R'\geq R$, 
and $R'\Gives \vc{V} \in \sem{R\Gives \Gamma}$.
By induction hypothesis, we know that 
$R'\Gives [\vc{V}/\vc{x}]M \in \sem{R\Gives (\regtype{r}{A},e_{1})}$ 
and $R'\Gives [\vc{V}/\vc{x}]N \in \sem{R\Gives (A,e_{2})}$.
Then for any $R''\geq R'$, 
$[\vc{V}/\vc{x}]M, \sem{R}(R'')$ normalizes to 
$r,\sem{R}(R'')$ and 
$[\vc{V}/\vc{x}]N, \sem{R}(R'')$ normalizes
to $V,\sem{R}(R'')$ where
$R''\Gives V\in \sem{R\Gives (A,\emptyset)}$.
Suppose $R=R_1,r:A,R_2$.
By definition, 
$\sem{R}(R'')(r)=\set{V' \mid R''\Gives V'\in \sem{R_{1}\Gives (A,\emptyset)}}$.
By proposition \ref{int-prop}(restriction), we know that
if $R''\Gives V\in \sem{R\Gives (A,\emptyset)}$ then 
$R''\Gives V \in  \sem{R_{1} \Gives (A,\emptyset)}$.
Therefore, $V \in \sem{R}(R'')(r)$, and 
the assignment normalizes to $*,\sem{R}(R'')(r)$.
It follows that 
$R''\Gives [\vc{V}/\vc{x}](\st{M}{N})$
belongs to $\sem{R\Gives (\Ter,e_1\union e_2\union  \set{r})}$.

\item  Suppose $R;\Gamma \Gives \gt{M} :(A,e\union \set{r})$ is
derived from $R;\Gamma \Gives M:(\regtype{r}{A},e)$.
Moreover, suppose $R'\geq R$, 
and $R'\Gives \vc{V} \in \sem{R\Gives \Gamma}$.
By induction hypothesis, we know that 
$R'\Gives [\vc{V}/\vc{x}]M \in \sem{R\Gives (\regtype{r}{A},e)}$.
Then for any $R''\geq R'$, $[\vc{V}/\vc{x}]M, \sem{R}(R'')$ normalizes to 
$r,\sem{R}(R'')$.
Thus $\gt{[\vc{V}/\vc{x}]M},\sem{R}(R'')$ will reduce to
$V,\sem{R}(R'')$ where $V\in \sem{R}(R'')(r)$ which is not
empty by proposition \ref{int-prop}(not-emptiness). 
Suppose $R=R_1,r:A,R_2$.
We know that $R''\Gives V \in \sem{R_1\Gives (A,\emptyset)}$
and by proposition \ref{int-prop}(extension) we 
conclude that
$R''\Gives V \in \sem{R\Gives (A,\emptyset)}$. \qed
\end{itemize}

\subsection{Proof of corollary \ref{cor-sound} (termination)}
\Proofitemf{(1)}
By definition, if $R\Gives M\in \sem{R\Gives (A,e)}$ then
$R; \Gives M:(A,e)$.
On the other hand, as a special case of theorem \ref{main-thm},
if $R; \Gives M:(A,e)$ is derivable
then $R\Gives M \in \sem{R\Gives (A,e)}$.

\Proofitem{(2)}
Suppose we have $R; \Gives M_1,\ldots,M_n:e$. Then
we have $R; M_i:(A_i,e_i)$ for $i=1,\ldots,n$.
By theorem \ref{main-thm}, the evaluation of 
$M_i,\sem{R}(R)$ is guaranteed to terminate in $V_i,\sem{R}(R)$,
for some value $V_i$. 
Now any reduction starting from $M_1,\ldots,M_n$ 
can be simulated step by step by a reduction of
$M_1,\ldots,M_n,\sem{R}(R)$ and therefore it must
terminate. \qed

\subsection{Decomposition for the timed/synchronous system}
Recall that $\Gives^{\w{ef}}$ denotes provability in the effect-free system.

\begin{proposition}[decomposition extended]\label{decomp-revisited}
If $\Gives^{\w{ef}} M: A$ is a well-typed closed thread 
then exactly one of the following situations arises where
$E$ is a time insensitive evaluation context:
(1) $M$ is a value;
(2) $M=E[\Delta]$ and $\Delta$ has the shape $(\lambda x.N)V$,
  $\st{r}{V}$, or $\gt{r}$; or
(3) $M=E[\pres{E'[\Delta]}{N}]$ and 
$\Delta$ has the shape $V$, $(\lambda x.N)V$, $\st{r}{V}$, or $\gt{r}$. 
\end{proposition}
\Proof
By induction on the structure of $M$.  
We consider in some detail the case for the else-next (cf.
proof \ref{proof-decomp} for other cases).

\Proofitem{\pres{M_1}{M_2}} 
We apply the inductive hypothesis to $M_1$, and we have three cases:
(1) $M_1$ is a value,
(2) $M_1=E_1[\Delta_1]$ with $E_1$ time insensitive, and
(3) $M_1=E_1[\pres{E_2[\Delta_1]}{N}]$ with $E_1$ time insensitive.
We note that in each case we fall in case 3 where the insensitive evaluation context
is $[~]$. \qed

\subsection{Basic properties for the timed/synchronous extensions}

\begin{proposition}[basic properties, stratified extended]\label{basic-prop-extended}
The following properties hold in the  stratified, timed/synchronous system.
\begin{description}

\item[weakening]
If $R;\Gamma \Gives M:(A,e)$ and $R,R' \Gives \Gamma,\Gamma'$ then
 $R,R';\Gamma,\Gamma' \Gives M:(A,e)$.

\item[substitution]
If $R;\Gamma,x:A \Gives M:(B,e)$ and 
$R;\Gamma \Gives N:(A,\emptyset)$ then 
$R;\Gamma \Gives [N/x]M:(B,e)$.

\item[context substitution]
If $R;\Gamma,x:A \Gives E[x]:(B,e)$ where $x$ is not free in the evaluation context $E$ and 
$R;\Gamma \Gives N:(A,e')$ then 
$R;\Gamma \Gives E[N]:(B,e\union e')$.

\item[subject reduction]
If $R,R';\Gamma \Gives \vc{M},S:(\Beh,e)$, 
$R \Gives e$, and $\vc{M},S \arrow \vc{M'},S'$  then
$R,R';\Gamma \Gives \vc{M'},S':(\Beh,e)$,
$S_{\mid \w{dom}(R')} = S'_{\mid \w{dom}(R')}$, and
$\vc{M},S_{\mid \w{dom}(R)} \arrow \vc{M'},S'_{\mid\w{dom}(R)}$.
Moreover, if $\vc{M}=M$ and $R,R'\Gives M:(A,e)$ then 
$\vc{M'}=M'$ and  $R,R';\Gives M':(A,e)$.

\item[tick reduction]
If $R; \Gives \vc{M},S:(\Beh,e)$, 
and $\vc{M},S \act{\tick} \vc{M'},S'$  then
$S=S'$ and there is an effect $e'$ such that
$R; \Gives \vc{M'},S:(\Beh,e')$.
\end{description}
\end{proposition}
\Proof
\begin{description}

\item[weakening/substitution]
The proofs of weakening and substitution proceed as in 
the proof \ref{proof-basic-prop}.

\item[context substitution]
We note that a proof of  $R;\Gamma \Gives M:(A,e)$ 
consists of a proof of $R;\Gamma \Gives M:(A',e')$,
where  $R\Gives (A',e')\leq (A,e)$, followed by  
a sequence of subtyping rules.
To prove context substitution, we proceed by induction
on the proof $R;\Gamma,x:A\Gives E[x]:(B,e)$ and by
case analysis on the shape of $E$.

\item[subject reduction]
To prove subject reduction, we start by noting that if
$R;x:A\Gives E[x]:(B,e)$ then $R;x:A\Gives \w{red}(E)[x]:(B,e)$.
In other terms, the elimination of the pending else-next branches from
the evaluation context preserves the typing.
Then we proceed by analysing the redexes as in proof \ref{proof-basic-prop}.

\item[tick reduction]
The interesting case is when 
$M=E[\pres{E'[\Delta]}{N}]$, 
$E$ is time insensitive, 
$\Delta$ has the shape $V$ or $\gt{r}$, and 
$M \act{\tick} E[N]$.
Suppose $R;\Gives M:(A,e)$.
Then the typing of the else-next guarantees that 
$R;\Gives E'[\Delta]:(B,e_1)$ and $R;\Gives N:(B,e_2)$
for some $B,e_1,e_2$ where $e_1$ and $e_2$ may be incomparable.
Then we can conclude $R;\Gives E[N]:(A,e')$ where 
the effect $e'$ is contained in $\w{dom}(R)$ but may 
be incomparable with $e$. \qed

\end{description}

\subsection{Proof of proposition \ref{else-next-elim} (simulation)}

\Proofitemf{(1)}
A straightforward induction on the typing.

\Proofitem{(2)} Immediate extension of step (1).

\Proofitem{(3)}  First we check that the translation commutes
with the substitution. Also, we extend the translation to
evaluation contexts, assuming $\tra{[~]}=[~]$, and check that
$\tra{E}$ is again an evaluation context.  
Then we proceed by case analysis on the reduction rule.

\Proofitem{(4)} Every reduction in $P$ corresponds to a reduction
in $\tra{P}$. \qed

\subsection{Proof of proposition \ref{derived-rule2-fix} (type
  fixed-point, revisited)}
The proof is a variation of the one for proposition \ref{derived-rule1-fix}.
Suppose $r:A\act{e}B \in R$ (hence $r\notin e$). 
Then $R; \Gives \lambda x.\gt{r}x: (A\act{e\union\set{r}}B,\emptyset)$.
By proposition \ref{basic-prop-extended}(substitution), 
$R;\Gamma\Gives M': (A\act{e}B,\emptyset)$ where
$M'=[\lambda x.\gt{r}x/f]M$.
From this we derive:
$R;\Gamma\Gives M'':(A\act{e}B,\set{r})$ where
$M''=\gt{\regterm{r}{\lambda x.M'x}}$. 
This judgement can be weakened to $R;\Gamma,x:A\Gives M'':(A\act{e}B,\set{r})$
which combined with $R;\Gamma,x:A\Gives x:(A,\emptyset)$ leads to
$R;\Gamma\Gives \lambda x.M''x:(A\act{e\union\set{r}}B,\emptyset)$ where
$\lambda x.M''x= \s{fix}_{r} f.M$, as required. \qed

\section{Summary of syntax, operational semantics, and typing rules}\label{summary-sec}
Table \ref{summary1} summarizes the main syntactic categories, the evaluation rules
for the computation within an instant (relation $\arrow$), and the rules
for the passage of time (relation $\act{\tick}$).
Table \ref{summary2} summarizes the typing rules for the 
unstratified and stratified systems
which differ just in the judgements for region contexts and types.

\begin{table}
{\footnotesize
\begin{center}
{\sc Syntactic categories}
\end{center}
\[
\begin{array}{ll}

x,y,\ldots         
&\mbox{(variables)} \\

r,s,\ldots         
&\mbox{(regions)} \\

e,e',\ldots
&\mbox{(finite sets of regions)} \\

A::= \Ter \Alt \regtype{r}{A} \Alt (A \act{e} A) \Alt (A\act{e} \Beh)
&\mbox{(types)} \\

\alpha::= A \Alt \Beh 
&\mbox{(types or behaviour)} \\

R::= r_1:A_1,\ldots,r_n:A_n 
&\mbox{(region context)} \\

\Gamma::= x_1:A_1,\ldots,x_n:A_n
&\mbox{(context)} \\

M::= x \Alt r \Alt * \Alt \lambda x.M \Alt MM \Alt \gt{M} \Alt \st{M}{M} \Alt  
\pres{M}{M} \Alt M,M &\mbox{(terms)} \\

V::= r \Alt * \Alt \lambda x.M
&\mbox{(values)} \\

v,v',\ldots
&\mbox{(sets of value)} \\
 
S::= \store{r}{v} \Alt S,S 
&\mbox{(stores)} \\

X::= M \Alt S
&\mbox{(stores or terms)}\\

P::= X \Alt X,P 
&\mbox{(programs)} \\

E::= [~] \Alt EM \Alt VE \Alt  \gt{E} \Alt \st{E}{M} \Alt \st{r}{E} \Alt \pres{E}{M}
&\mbox{(evaluation contexts)}

\end{array}
\]
\begin{center}
{\sc Evaluation rules within an instant}
\end{center}
\[
\begin{array}{c}

\infer{}{E[(\lambda x.M)V] \arrow \w{red}(E)[[V/x]M]} 
\qquad
\infer{}{E[\gt{r}],\store{r}{V} \arrow \w{red}(E)[V],\store{r}{V}} \\ \\ 

\infer{}{E[\st{r}{V}]\arrow \w{red}(E)[*],\store{r}{V}}

\qquad \infer{P\arrow P'}
{P,P'' \arrow P',P''}

\end{array}
\]
\begin{center}
{\sc Rules for the passage of time}
\end{center}
\[
\begin{array}{c}

\infer{}
{V\act{\tick} V}

\qquad
\infer{M=E[\gt{r}]\quad E \mbox{ time insensitive}}
{M \act{\tick} M}
\\ \\

\infer{M=E[\pres{E'[\Delta]}{N}] \quad  E \mbox{ time insensitive}\quad \Delta::=V\Alt \gt{r}}
{M \act{\tick} E[N]} \\ \\

\infer{}
{S\act{\tick} S}

\qquad \infer{P_1,P_2\not\arrow \quad P_i \act{\tick} P'_i\quad i=1,2}
{P_1,P_2 \act{\tick} P'_{1},P'_{2}}

\end{array}
\]}
\caption{Syntactic categories and operational semantics}\label{summary1}
\end{table}

\begin{table}
{\footnotesize
\[
\begin{array}{c}

\mbox{{\sc Unstratified region contexts and types}}  \\

\infer{}
{R\dcl \Ter}
\qquad

\infer{}
{R\dcl \Beh}
\qquad

\infer{R\dcl A\quad R\dcl \alpha \quad e\subseteq \w{dom}(R)}
{R\dcl (A\act{e} \alpha)}

\qquad

\infer{r:A\in R}
{R\dcl \regtype{r}{A}} \\ \\ 

\infer{\qqs{r\in \w{dom}(R)}{R\dcl R(r)}}
{R\Gives}

\qquad

\infer{R\Gives\quad R\dcl \alpha}
{R\Gives \alpha} 

\qquad

\infer{R\Gives \alpha \quad e\subseteq \w{dom}(R)}
{R\Gives (\alpha,e)} 
\\ \\

\mbox{{\sc Stratified region contexts and types}}  \\

\infer{}{\emptyset \Gives}

\qquad

\infer{R\Gives A\quad r\notin \w{dom}(R)}
{R,r:A\Gives} 
\qquad

\infer{R\Gives}
{R\Gives \Ter} 

\qquad

\infer{R\Gives}
{R\Gives \Beh}
\\ \\

\infer{R\Gives \quad r:A \in R}
{R\Gives \regtype{r}{A}} 

\qquad

\infer{R\Gives A\quad R\Gives \alpha \quad e\subseteq \w{dom}(R)}
{R\Gives (A\act{e} \alpha)n} 

\qquad

\infer{R\Gives \alpha \quad e\subseteq\w{dom}(R)}
{R\Gives (\alpha,e)} \\  \\

\mbox{{\sc Subtyping rules}} \\

\infer{R\Gives \alpha}{R\Gives \alpha \leq \alpha}

\qquad

\infer{\begin{array}{c}
R\Gives A'\leq A\quad  R\Gives \alpha \leq \alpha'\\
e\subseteq e' \subseteq \w{dom}(R) 
\end{array}}
{R\Gives (A\act{e} \alpha) \leq (A' \act{e'} \alpha')} 

\qquad

\infer{\begin{array}{c}
R\Gives \alpha \leq \alpha'\\ 
e\subseteq e' \subseteq \w{dom}(R)
\end{array}}
{R\Gives (\alpha,e) \leq (\alpha',e')} \\ \\

\mbox{{\sc Terms, stores, and programs}} \\ 

\infer{R\Gives \Gamma \quad x:A\in \Gamma}
{R;\Gamma\Gives x:(A,\emptyset)}

\qquad 
\infer{R\Gives \Gamma \quad r:A\in R}
{R;\Gamma\Gives r:(\regtype{r}{A},\emptyset)} 

\qquad
\infer{R\Gives \Gamma}
{R;\Gamma \Gives *: (\Ter,\emptyset)} \\ \\

\infer{R;\Gamma,x:A \Gives M: (\alpha,e)}
{R;\Gamma\Gives \lambda x.M :(A\act{e} \alpha,\emptyset)}

\qquad

\infer{R;\Gamma \Gives M: (A\act{e_{2}} \alpha ,e_{1}) \quad R;\Gamma\Gives N: (A,e_{3})}
{R;\Gamma \Gives MN :(\alpha,e_1\union e_2\union e_3)} \\ \\

\infer{R;\Gamma \Gives M: (\regtype{r}{A}, e)}
{R;\Gamma \Gives \gt{M} : (A,e\union \set{r})} 

\qquad

\infer{R;\Gamma \Gives M: (\regtype{r}{A}, e_{1})\quad R;\Gamma \Gives N:(A,e_{2})}
{R;\Gamma \Gives \st{M}{N} : (\Ter,e_1\union e_2 \union \set{r})} \\ \\ 

\infer{R;\Gamma\Gives M:(A,e)\quad R;\Gamma \Gives N:(A,e')}
{R;\Gamma \Gives \pres{M}{N} : (A,e)}

\qquad
\infer{R;\Gamma \Gives M:(\alpha,e) \quad R\Gives (\alpha,e)\leq (\alpha',e')}
{R;\Gamma \Gives M:(\alpha',e')}  \\ \\ 

\infer{r:A \in R \quad \qqs{V\in v}{R;\Gamma \Gives V: (A,\emptyset)}}
{R;\Gamma \Gives \store{r}{v} :(\Beh,\emptyset)}

\qquad 

\infer{R;\Gamma \Gives X_i:(\alpha_i,e_i)\quad i=1,\ldots,n\geq 1}
{R;\Gamma \Gives X_1,\ldots,X_n:(\Beh,e_1\union \cdots\union e_n)}

\end{array}
\]}
\caption{Typing systems}\label{summary2}
\end{table}


\begin{thebibliography}{99}

\bibitem{AC98}
R.~Amadio and P.-L.~Curien.
\newblock Domains and Lambda Calculi.
\newblock {\em Cambridge University Press}.

\bibitem{AD07}
R.~Amadio and F.~Dabrowski.
\newblock Feasible reactivity in a synchronous $\pi$-calculus.
\newblock In Proc. ACM Principles and Practice of Declarative Programming.
pp 221-230, 2007.

\bibitem{Boudol07} 
G.~Boudol.
\newblock Typing termination in a higher-order concurrent imperative 
language.
\newblock In Proc. CONCUR, Springer LNCS 4703:272-286, 2007.


\bibitem{BG92} G.~Berry and G.~Gonthier. \newblock The Esterel
  synchronous programming language.  \newblock {\em Science of
    computer programming}, 19(2):87--152, 1992.

\bibitem{DS06} Y.~Deng and D.~Sangiorgi. 
\newblock Ensuring termination by typability.
\newblock {\em Information and Computation}, 
204(7):1045-1082, 2006.

\bibitem{Gal90} 
J.~Gallier. \newblock On Girard's {\em Candidats de Reductibilit\'{e}}. 
\newblock In {\em Logic and Computer Science}, Odifreddi (ed.), 
Academic Press, 123-203, 1990.

\bibitem{HR95}
M.~Hennessy, T.~Regan. 
\newblock A process algebra of timed systems. 
\newblock {\em Information and Computation}, 117(2):221-239, 1995.


\bibitem{LG88}
J.~Lucassen and D.~Gifford.
\newblock Polymorphic effect systems.
\newblock In Proc. ACM-POPL, 1988.

\bibitem{S06}
D.~Sangiorgi.
\newblock Termination of processes.
\newblock {\em Math. Struct. in Comp. Sci.},
16:1-39, 2006.

\bibitem{TT97}
M.~Tofte and J.-P.~Talpin.
\newblock Region-based memory management. 
\newblock {\em Information and Computation}, 132(2): 109-176, 1997.

\bibitem{YBH04} 
N.~Yoshida, M.~Berger, and K.~Honda.
\newblock Strong normalisation in the $\pi$-calculus. 
\newblock {\em Information and Computation}, 191(2):145-202, 2004.
\end{thebibliography}
\end{document}